\documentclass[10pt]{iopart}
\usepackage{graphicx}
\usepackage{subfig}
\usepackage{float}
\usepackage{epstopdf}
\usepackage{amsfonts}
\usepackage{color}
\usepackage{textpos}

\begin{document}

\begin{textblock*}{100mm}(\textwidth,-2cm)
\textcolor{blue}{\large Manuscript}
\end{textblock*}

\def\look#1{\textcolor{red}{#1}}

\title{Geometric effects in the electronic  transport of deformed nanotubes}

\author{Fernando Santos
\footnote{On leave from: Departamento de Matem\'atica  Universidade Federal 
de Pernambuco, 50670-901, Recife, PE, Brazil}}
\address{Wolfson Centre for Mathematical Biology, Mathematical Institute, University of Oxford, OX1 3LB Oxford, U.K.}
\author{S\'ebastien Fumeron, Bertrand Berche}
\address{ Statistical Physics Group, IJL, UMR Universit\'e de Lorraine - CNRS 7198
 BP 70239,  54506 Vand\oe uvre les Nancy, France}
 \author{Fernando Moraes
\footnote{On leave from: Departamento de F\'\i sica, CCEN, Universidade Federal da Para\'\i ba, Caixa Postal 5008, 58051-900, Jo\~ao Pessoa, PB, Brazil }}

\address{Departamento de F\'\i sica and Departamento de Matem\'atica, Universidade
Federal de Pernambuco, 
 50670-901 Recife, PE, Brazil}
 
\date{\today}

\begin{abstract}
Quasi-two-dimensional systems may exibit curvature, which adds three-dimensional influence to their internal properties. As shown by da Costa \cite{dacosta}, charged particles moving on a curved surface experience a curvature-dependent potential which greatly influence their dynamics. In this paper, we study the electronic ballistic transport in deformed nanotubes. The one-electron Schr\"odinger equation with open boundary conditions is solved numerically with a flexible MAPLE code made available as Supplementary Data. We find that the curvature of the deformations have indeed strong effects on the electron dynamics suggesting its use in the design of nanotube-based electronic devices.
\end{abstract}

\noindent{\it Keywords\/}: Nanotubes, Ballistic transport, Geometric effects


\maketitle
\ioptwocol
\section{Introduction}

Corrugated is graphene's  natural state \cite{novoselov}. The curvature associated to it leads to interesting phenomena such as curvature induced p-n junctions, band-gap opening and decoherence \cite{PhysRevB.81.205409}. Carbon nanotubes, on the other hand, are perfect cylindrical surfaces which can be deformed into wavy nanotubes from different techniques: axial compression \cite{PhysRevLett.84.1712}, combination of defect formation and electromigration \cite{doi:10.1021/nl061671j} or filling the nanotubes with fullerenes (``peapods'' \cite{smith}). 
As curvature influences the dynamics of quantum particles, it can be used to manipulate the electronic properties of two-dimensional materials \cite{PhysRevB.81.205409,:/content/aip/journal/jcp/142/19/10.1063/1.4921310} which can be measured.
Experimental characterization of the electronic properties of imperfect nanotubes can be achieved via different techniques. The conventional approach requires making contacts and measuring transport properties of the device~\cite{doi:10.1021/nl061671j}. Other possible techniques are based on near field microscopy, either dielectric force microscopy~\cite{NanoRes7-1623} or scanning tunneling microscopy~\cite{Ouyang27042001} which probe locally the electronic structure and properties without need of making contacts.

Ballistic electron transport in nanotubes is known to be drastically affected by variations of the tube geometry \cite{Marchi,Taira20075270} and therefore, of the local curvature. This effect is the main concern of the present article.
Of course, carbon nanotudes are one-dimensional structures which can exhibit metallic, semimetallic, or insulating properties depending on their chirality, i.e. the way they are mapped from a graphene sheet~\cite{Ouyang27042001,PhysRevLett.78.1932,PhysRevLett.79.5082}. Generically, the most conducting are the ``armchair'' carbon nanotubes. Their behaviour is described by a Luttinger liquid rather than a Fermi liquid, the shorter the nanotube cylinder, the stronger the Luttinger behaviour. This has been shown experimentally, see, e.g. \cite{nature397-598}.
 In spite of this, short-range electron-electron interactions have only weak effects and the deviations from the behaviour of non-interacting electrons occur only at very low temperature and for nanotubes of very small transverse size~\cite{PhysRevLett.79.5086}. 

Our aim in this paper is not to make a quantitatively realistic description of deformed nanotubes, but to use a simple independent electron approach in order to access qualitatively the ballistic transport regime, and possibly, arouse further studies on potential applications. With the examples reported here we intend to draw attention to the possibilities geometry can bring to the design of nanotubes with specific electronic properties. We
 model geometrically the deformed cylindrical surface of a nanotube in the following situations: shrunk nanotube, nanotube with a bump and a wavy peapod. We solve numerically the Schr\"odinger equation for a single particle moving in the deformed cylinder, taking into account the modification to the kinetic energy caused by the non-Euclidean surface and
the geometric potential induced by curvature. We then compute the transmittance as a function of the injection energy  for all cases. 
We developed a quite flexible MAPLE code (see Supplementary Data for the code and its consistency check) to solve the Schr\"odinger equation with open boundary conditions for a family of cylindrical symmetry surfaces, including the cylindrical junctions in Ref.~\cite{Marchi}.

\section{Quantum particle in a curved surface with azimuthal symmetry}
The problem of a free quantum particle moving in a  
curved surface was solved in a very realistic setting by R. C. T. da Costa \cite{dacosta} in a seminal paper published in 1981. The da Costa approach  has been applied to a wide range of two-dimensional systems, like rolled-up nanotubes \cite{PhysRevLett.113.227205}, thin magnetic shells \cite{PhysRevLett.112.257203} or spin transport on curved systems \cite{doi:10.1142/S2010324713400067}. In particular, the approach has been much used to study carbon-based systems like nanotubes and other curved forms of graphene \cite{0953-8984-23-17-175301}. The experimental verification of the geometric effects predicted by da Costa in a real physical system was done in \cite{0295-5075-98-2-27001} by measuring the high-resolution ultraviolet photoemission spectra of a C$_{60}$ peanut-shaped polymer. 

Considering a finite thickness $d$ for the ``surface" and a confining potential given by an infinite square well in the normal direction, da Costa found that the Schr\"odinger operator for a free particle in this confined geometry, in the limit $d\rightarrow 0$, is \cite{dacosta}
\begin{equation}
H = -\frac{\hbar^2}{2m}\triangle_{LB} + V_{\rm geo}, \label{HdaCosta}
\end{equation} 
where the Laplace-Beltrami operator is obtained from the curvilinear coordinates $x^i$ intrinsic to the surface and the metric tensor $g_{ij}$ ($g$ its determinant)
\begin{equation}
\triangle_{LB} = \sum_{i,j}\frac{1}{\sqrt{|g|}}\frac{\partial}{\partial x^i}\left(\sqrt{|g|}g^{ij}\frac{\partial}{\partial x^j}  \right), \label{LB}
\end{equation}
and the geometric potential is given by
\begin{equation}
V_{\rm geo}=- \frac{\hbar^2}{2m}\left(M^2 - K \right). \label{geompotential}
\end{equation}
Here, $M$ and $K$ refer to the mean and Gaussian curvatures, respectively. The geometric potential is a direct consequence of the quantization of the motion normal to the surface. 

Let us now focus on surfaces of revolution since our interest is to study corrugated nanotubes. A surface of revolution is obtained by rotation of a plane curve around an axis. The parametric equations for the surface of revolution can be written as $x= \rho(q) \cos\phi,\;\;y = q,\;\;z = \rho(q) \sin\phi$, where $q \in \mathbb{R}^1$ and $\phi \in \mathbb{S}^1$. With such parametrization, the first and second fundamental forms are respectively:
\begin{equation}
g_{ij}=\left(
\begin{array}{cc}
\rho(q)^2 & 0 \\
0 & 1+\rho^\prime(q)^2 \\    
\end{array} \right) \label{g}
\end{equation}
and
\begin{equation}
h_{ij}= \frac{1}{\sqrt{1+\rho^\prime(q)^2}}\left(
\begin{array}{cc}
\rho(q) & 0 \\
0 & -\rho^{\prime\prime}(q) \\   
\end{array} \right) .
\end{equation}

The mean and Gaussian curvatures are given respectively by $M=\frac{1}{2g}(g_{11}h_{22}+g_{22}h_{11}-2g_{12}h_{12})$ and $K=\frac{1}{g}(h_{11}h_{22}-h_{12}h_{21})$, so that the geometric potential is analytically obtained in terms the nanotube parametrization $\rho(q)$:
\begin{equation}
V_{\rm geo}=-\frac{\hbar^2}{8m}\frac{[1+\rho^\prime(q)^2+\rho(q)\rho^{\prime\prime}(q)]^2}{\rho(q)^2[1+\rho^\prime(q)^2]^3}. \label{V}
\end{equation}

Therefore, from Eqs. (\ref{HdaCosta}) and (\ref{LB}),  the Hamiltonian follows
\begin{equation}
H= -\frac{\hbar^2}{2m}\left[(g_{11})^{-1}\frac{\partial^2}{\partial \phi^2}
 +\frac{1}{\sqrt{g}}\frac{\partial}{\partial q} \left( \sqrt{g}(g_{22})^{-1}   \frac{\partial}{\partial q}\right)\right]  \nonumber  + V_{\rm geo} .
\end{equation}

The potential $V_{\rm geo}$ depends only on the $q$ coordinate, so we can separate the Schr\"odinger equation into two 1D equations
$\Phi^{\prime\prime} + \ell^2\Phi = 0$ ,
with $\ell \in \mathbb{Z}$ in order to satisfy angular periodicity and
\begin{eqnarray}
\Psi^{\prime\prime} + F(q)\Psi^{\prime}+ G(q)\left[\frac{2m}{\hbar^2} (E - V_{\rm geo})] - \frac{\ell^2}{\rho^2}\right]\Psi  = 0 .
 \label{q}
\end{eqnarray}
Here, $E$ is the total energy, $\Phi=Ae^{i\ell\phi}$ are the eigenfunctions of the angular momentum $\ell \hbar$ along the  $y$-axis and $\Psi(q)$, the longitudinal eigenfunction. The functions $F$ and $G$ in Eq.~(\ref{q}) are
\begin{equation}
F(q)= \frac{\rho^\prime}{\rho}\left[1-\frac{\rho\rho^{\prime\prime}}{1+(\rho^\prime)^2}\right],
\quad
G(q)= 1+(\rho^\prime)^2 .
\end{equation}

We model the corrugated nanotube as two  semi-infinite cylinders of radius $R$, joined by a surface of revolution generated by a curve $\rho(y)$ in the range $0<y<L$, such that $\rho(0)=\rho(L)=R_1$. In the region $y<0$, total energy is the sum of the injection energy $E_k=\hbar^2 k_{0}^2/(2m^{*})$ and the angular part $E_\ell=\hbar^2\ell^2/(2m^*R_{1}^2)$. The mass $m$ of the particle is replaced by its effective mass $m^* = 0.173 m_e$ in order to facilitate comparison with reference \cite{Marchi}.

In the next section we solve Eq. (\ref{q}) with open (Robin) boundary conditions using the quantum transmitting boundary method \cite{lent}. With these techniques we built a numerical code which we validated by reproducing the result for the electron transmittance in an axially symmetric cylindrical junction studied in \cite{Marchi}. 

\section{Methodology}
Recalling that $\rho=\rho(y)$, we see that Eq. (\ref{q}) is of the form
\begin{equation}
\Psi(y)^{\prime\prime} + V_1(y) \Psi(y)^{\prime}  + V_2(y) \Psi(y) = 0. \label{Psieq}
\end{equation}
By making $\Psi(y)=\varphi(y)\lambda(y)$ we get
\begin{equation} \label{int-eq}
\varphi^{\prime\prime} + \left(2\frac{\lambda^{\prime}}{\lambda} + V_1 \right) \varphi^{\prime}  + \left(\frac{\lambda^{\prime\prime}}{\lambda}+ V_1 \frac{\lambda^{\prime}}{\lambda} +V_2 \right) \varphi = 0.
\end{equation}
Now, for $2\frac{\lambda^{\prime}}{\lambda} + V_1 =0$, one gets $\lambda(y) = e^{-\frac{1}{2}P(y)}$, where $P(y)$ is the primitive for the function $V_1(y)$. Therefore, (\ref{int-eq}) becomes
\begin{equation}
\varphi^{\prime\prime}(y)  + \left( -\frac{1}{4} V_1^2 - \frac{1}{2} V_1^{\prime} + V_2 \right) \varphi = 0\label{phieq}
\end{equation}
and transforming back to $\Psi$:
$\Psi(y) = e^{-\frac{1}{2}P(y)} \varphi(y)$.

Since we are interested in the transmittance  due to the deformation of the tube, we consider the injection of electrons of energy $E_k$ coming from the negative part of the $y$-axis. Thus, we have
\begin{eqnarray}
\Psi (y) & = & a_0 e^{ik_0 y} + b_0 e^{-i k_0 y} \,\,\, \mathrm{for} \, y \leq 0,\ \nonumber \\ 
            & = &a_L e^{-ik_L (y-L)} + b_L e^{i k_L( y-L)} \,\,\,\mathrm{for}\, y \geq L, \label{psiinout}
\end{eqnarray}
where 
\begin{eqnarray}
k_0 &=& \sqrt{\frac{2m^{*}}{\hbar^2}(E_k-V(0))} \label{k0}\\
k_L&=& \sqrt{\frac{2m^{*}}{\hbar^2}(E_k-V(L))} \label{kL}
\end{eqnarray}
are the incident and transmitted electron wavectors, respectively.  By doing this we are assuming that there are perfect matching contacts at both ends of the tube. Since our aim in this article is to present a way of designing specific electronic properties by manipulation of the nanotube shape we decided, following Ref. \cite{Marchi}, to use this simpler approach. A complete treatment of the boundary conditions including the input and output leads can be found in \cite{lent}.

Notice that  $V_1(y)=0$ for $y$ not in the range $0<y<L$, which makes Eqs. (\ref{Psieq}) and (\ref{phieq}) identical. So, Eq. (\ref{psiinout}) is also valid for $\varphi(y)$ and then $\Psi(0)=\varphi(0)$ as well as $\Psi(L)=\varphi(L)$. As we will see below this implies that the reflectance and transmittance do not depend on $\lambda(y)$ and can be obtained from $\varphi(y)$ directly. We choose the normalization of the incident wavefunction such that $a_0=1$. Also, considering only outgoing waves, in the $y\geq 0$ region we have  $a_L =0$.

For the above normalization, the transmittance  may be obtained from the probability current density
\begin{equation}
j=\frac{\hbar}{2mi}\left( \Psi^* \Psi^{'}- \Psi \Psi^{* '}\right)
\end{equation}
such that the incident current is $j_{inc}= \frac{\hbar k_0}{m^{*}}$, the reflected current is $j_{ref} = \frac{\hbar k_0}{m^{*}} |b_{0}|^2$ and the transmitted current $j_{trans}= \frac{\hbar k_L}{m^{*}} |b_L|^2$. Hence, the transmittance is
$T=\frac{j_{trans}}{j_{inc}}=\frac{k_L}{k_0}| b_L |^2$ 
and the reflectance,
$R=\frac{j_{ref}}{j_{inc}}=| b_0 |^2 $.

Using the boundary conditions $a_0=1$ and $a_L =0$, it comes after some algebra that 
\begin{eqnarray}
T=\frac{k_L}{k_0} \left| \varphi(L) \right|^2,  \label{transnew}\\
R=|\varphi(0)-1 |^2 \label{refnew} .
\end{eqnarray}
Then, the problem reduces to finding $\varphi(0)$ and $\varphi(L)$ by solving the coupled differential and algebraic equations (Eq. (\ref{phieq}) and boundary conditions) in the range $0\leq y\leq L$. This is the essence of the open boundary condition method for solving ordinary differential equations with Robin boundary conditions. With the above expressions we implemented a MAPLE code 
to find, for each injection energy, $\varphi(0)$  and $\varphi(L)$ and, consequently, the transmittance and the reflectance as specified by Eqs. (\ref{transnew}) and (\ref{refnew}), respectively. 
In order to input the energy in ${\rm meV}$ and distances in ${\rm nm}$ we use a mixed units system where the electron mass is $m_e = 5.68 \times 10^{-27}  {\rm meV.s}^2/{\rm nm}^2$ and Planck's constant is $\hbar= 6.58 \times 10^{-13} {\rm meV.s}$  .

\section{Results}
Resonance peaks correspond to quasibound states. These  are  states associated to a quantum well where a particle is primarily confined but has a finite probability of tunnelling out  and escaping. In the nanotube, the geometric potential, if deep enough, may have such states. Although most of the incident electrons are reflected back by the  geometric potential, if the energy of the electron coincides with that of a quasibound state this makes it easier for it to tunnel to the inner region of the potential and thus tunnel out of it on the opposite side.  If the  potential becomes deeper, the energy levels of the quasibound states shift downward implying a shift of the resonant peaks to lower energies. This is in fact what is seen in the results described below.

Looking at Eq. (\ref{q}) we see that for angular momentum $\ell\neq 0$ a repulsive term $\frac{\ell^2}{\rho^2}$ is added to the geometric potential ({\ref{V}). So, the effect of the centrifugal term is to make the potential well shallower and consequently reducing the number of quasibound states, therefore of the resonances in the transmittance.  For this reason, in what follows we consider only  the cases of zero angular momentum ($\ell=0$) which gives the general physical picture of the system.

\begin{figure*}[ht]
\centering
    \subfloat{\includegraphics[width=0.3\linewidth]{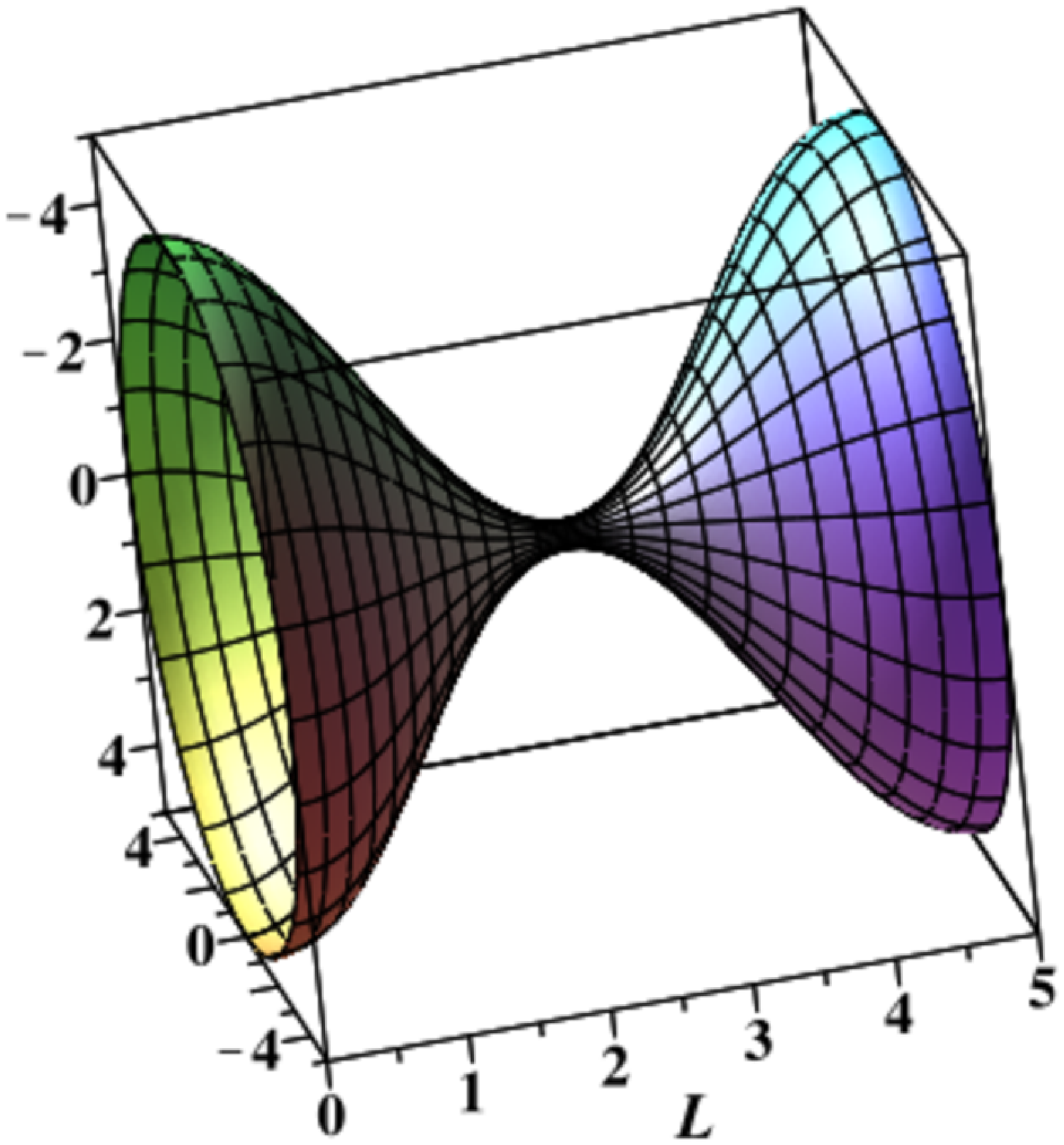}}            
    \subfloat{\includegraphics[width=0.36\linewidth]{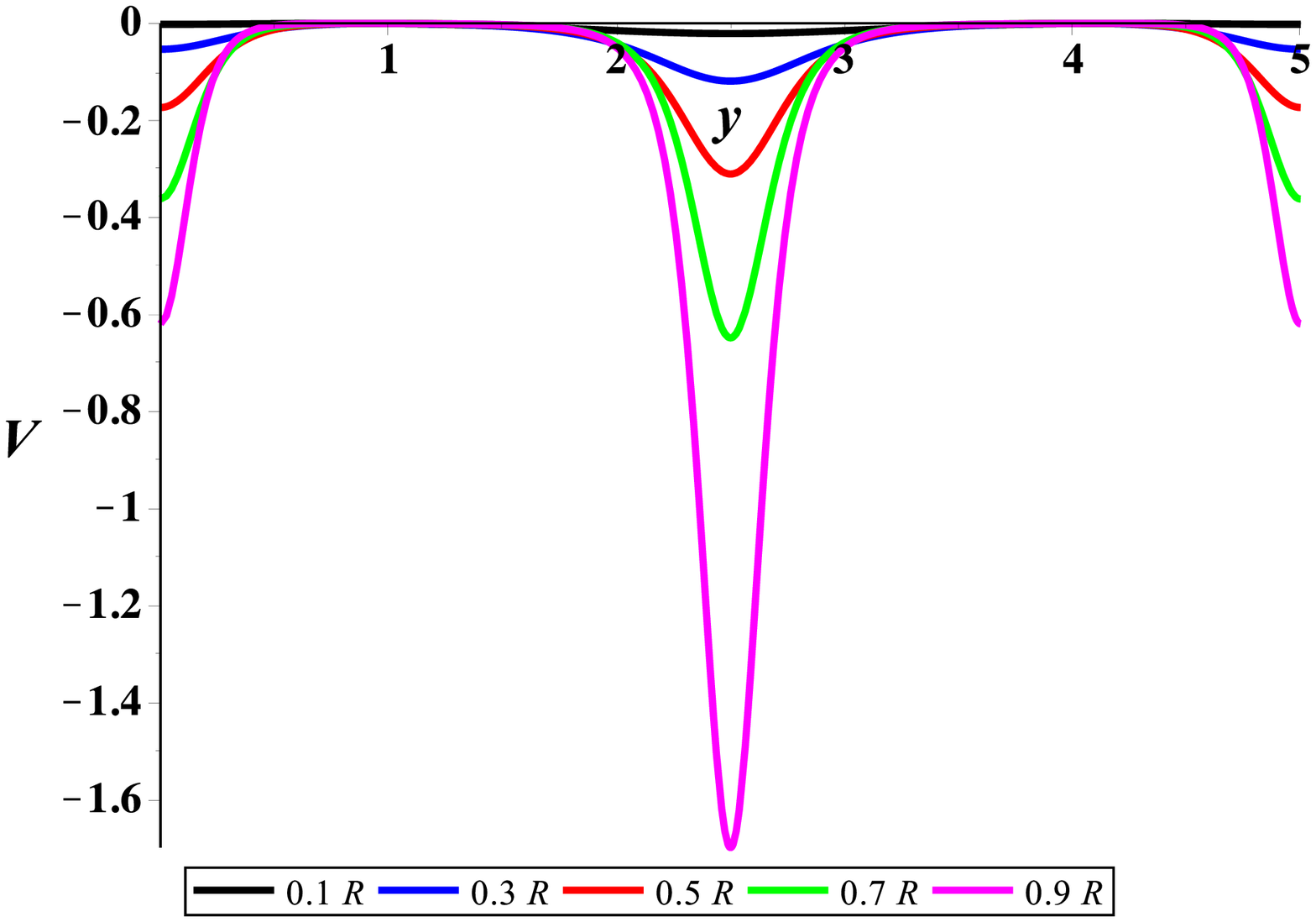}}
    \subfloat{\includegraphics[width=0.31\linewidth]{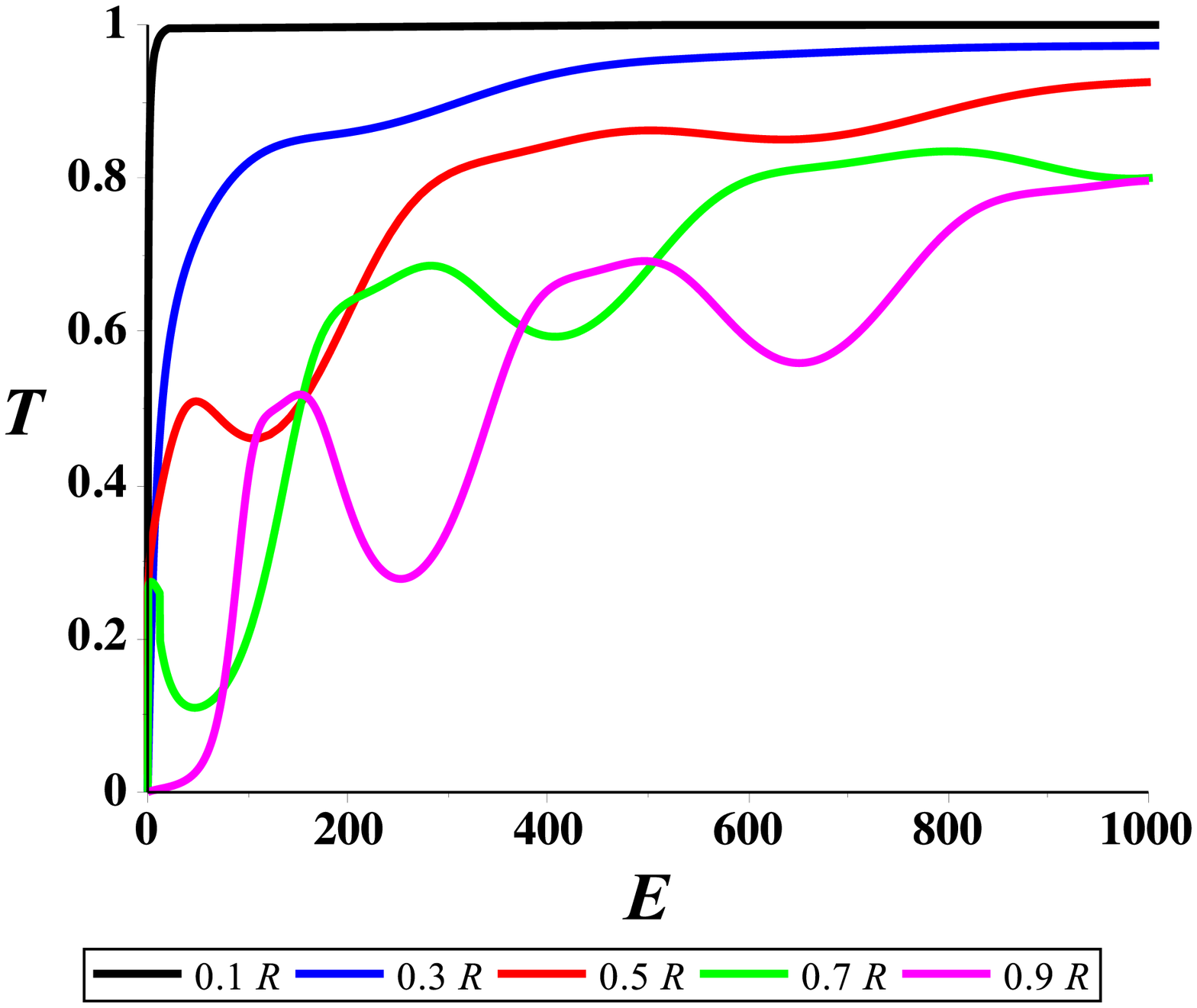}}
          \caption{ (a) Pinched nanotube. (b) Geometric potential due to the deformation shown in (a) and (c) transmittance as function of incident energy (in meV) for different waist sizes ($\epsilon R$).} \label{Fig1}
\centering
    \subfloat{\includegraphics[width=0.3\linewidth]{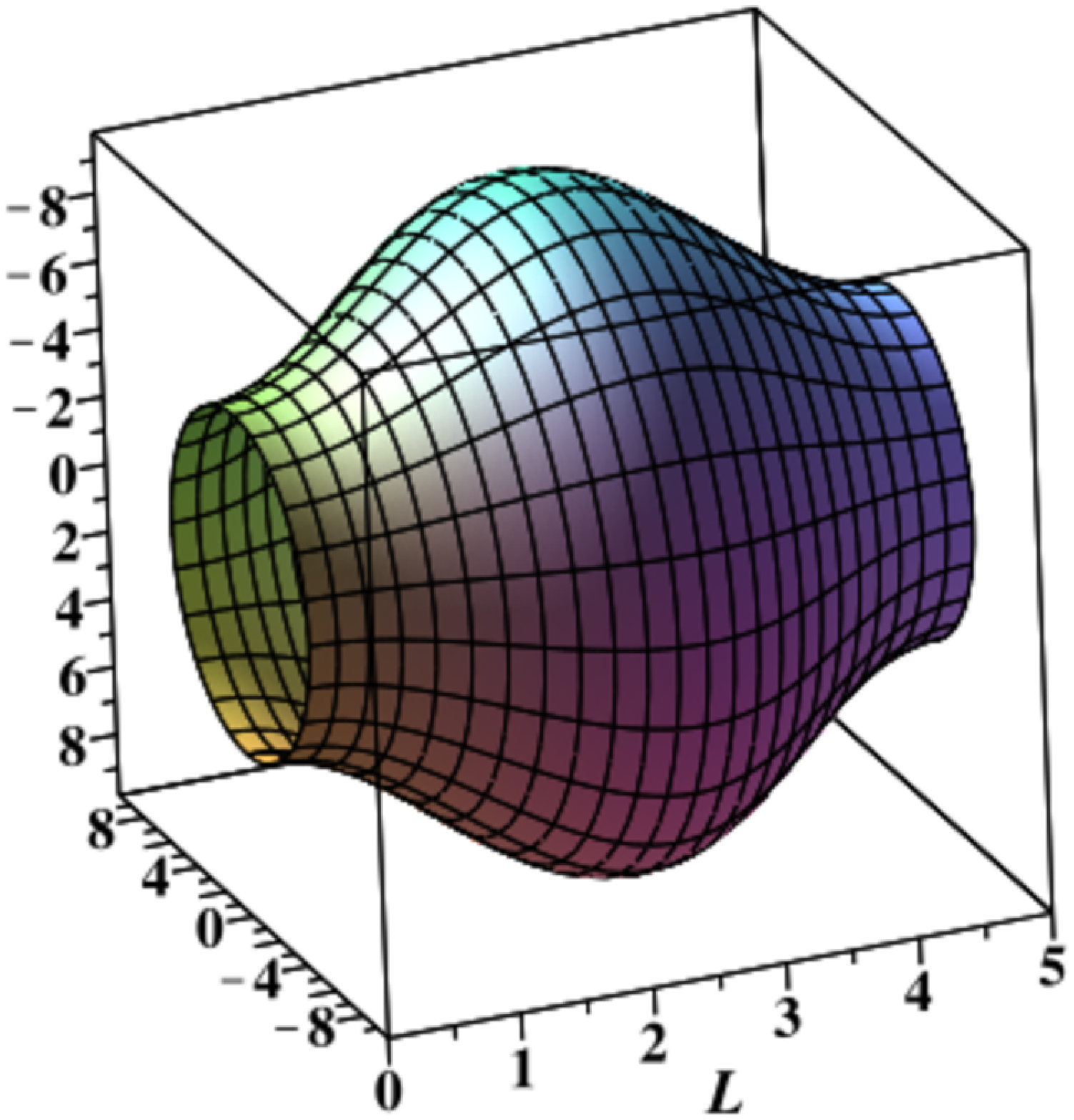}}
    \subfloat{\includegraphics[width=0.36\linewidth]{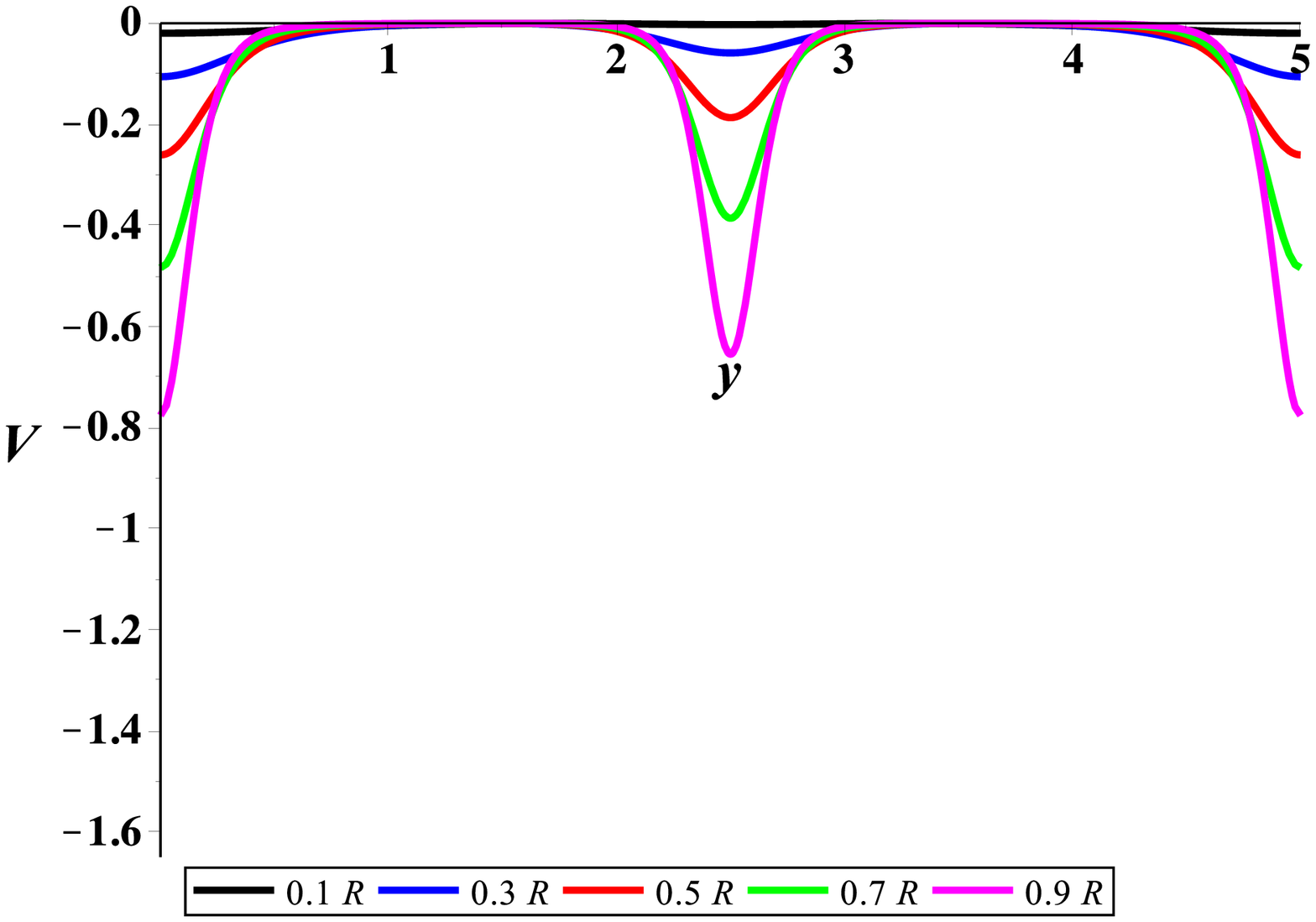}}
    \subfloat{\includegraphics[width=0.32\linewidth]{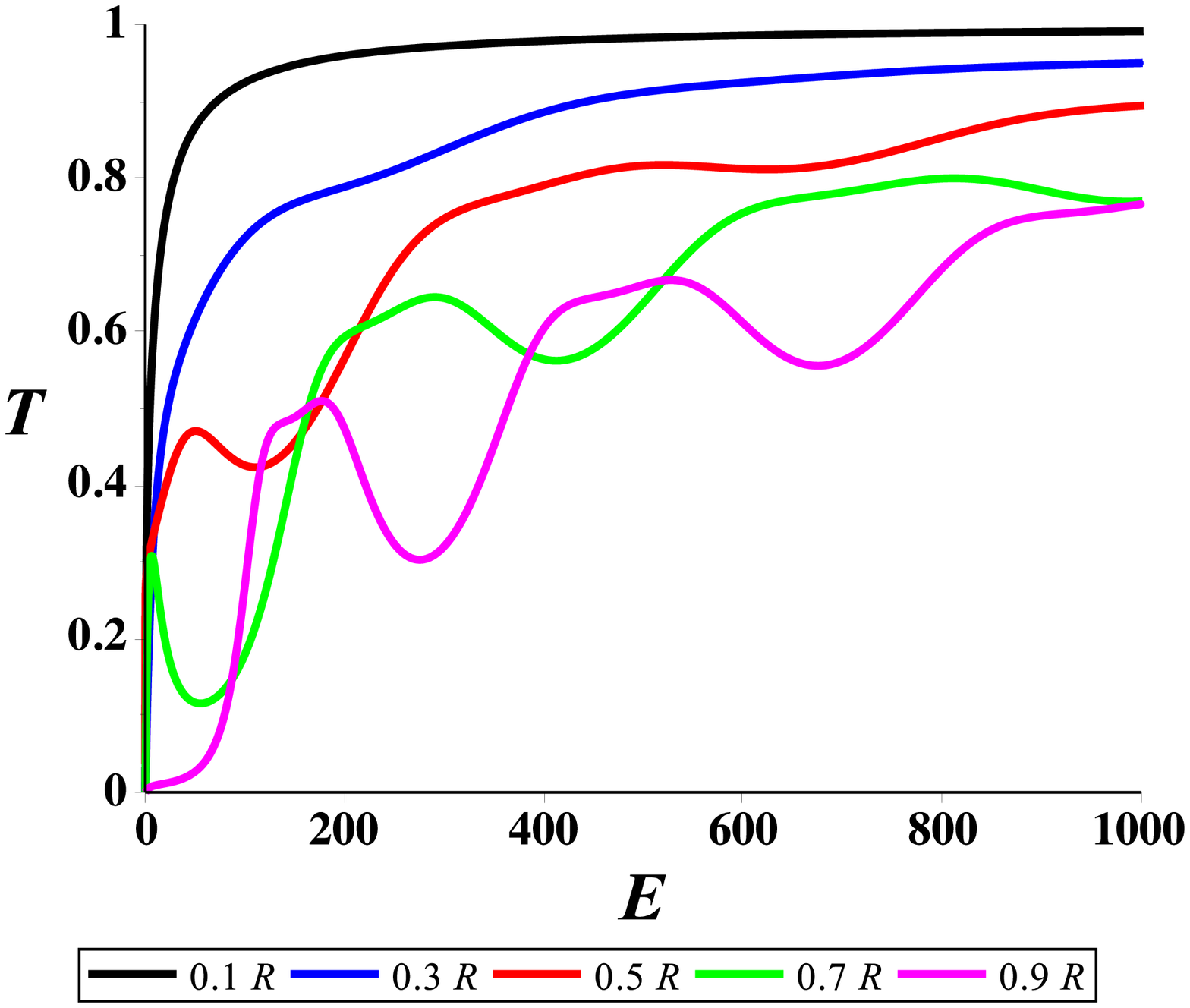}}
          \caption{ (a) Nanotube with a single bump. (b) Geometric potential due to the deformation shown in (a) and (c) transmittance as function of incident energy (in meV) for different bump sizes ($\epsilon R$).} \label{Fig2}
\centering
    \subfloat{\includegraphics[width=0.30\linewidth]{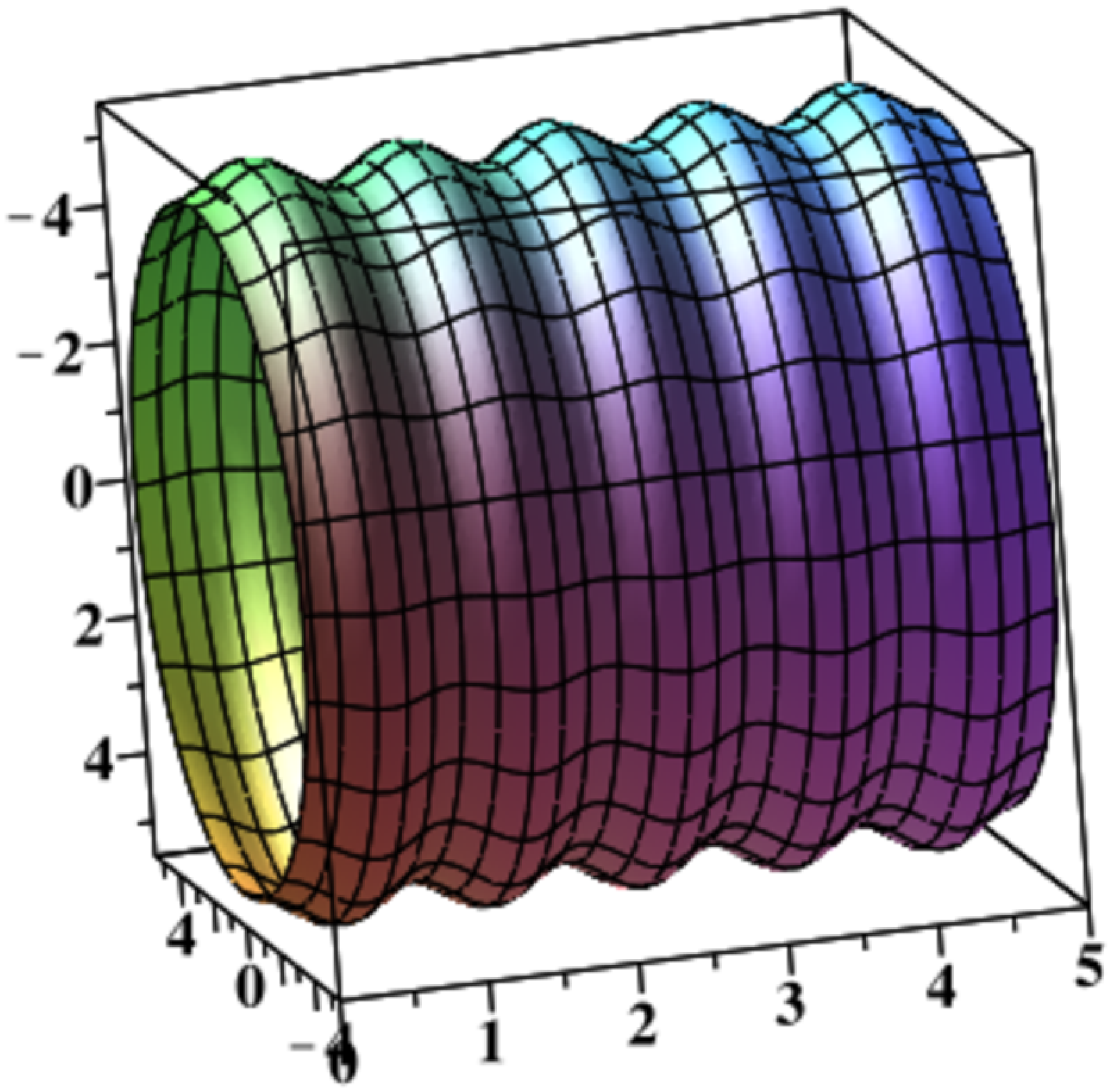}}
    \subfloat{\includegraphics[width=0.36\linewidth]{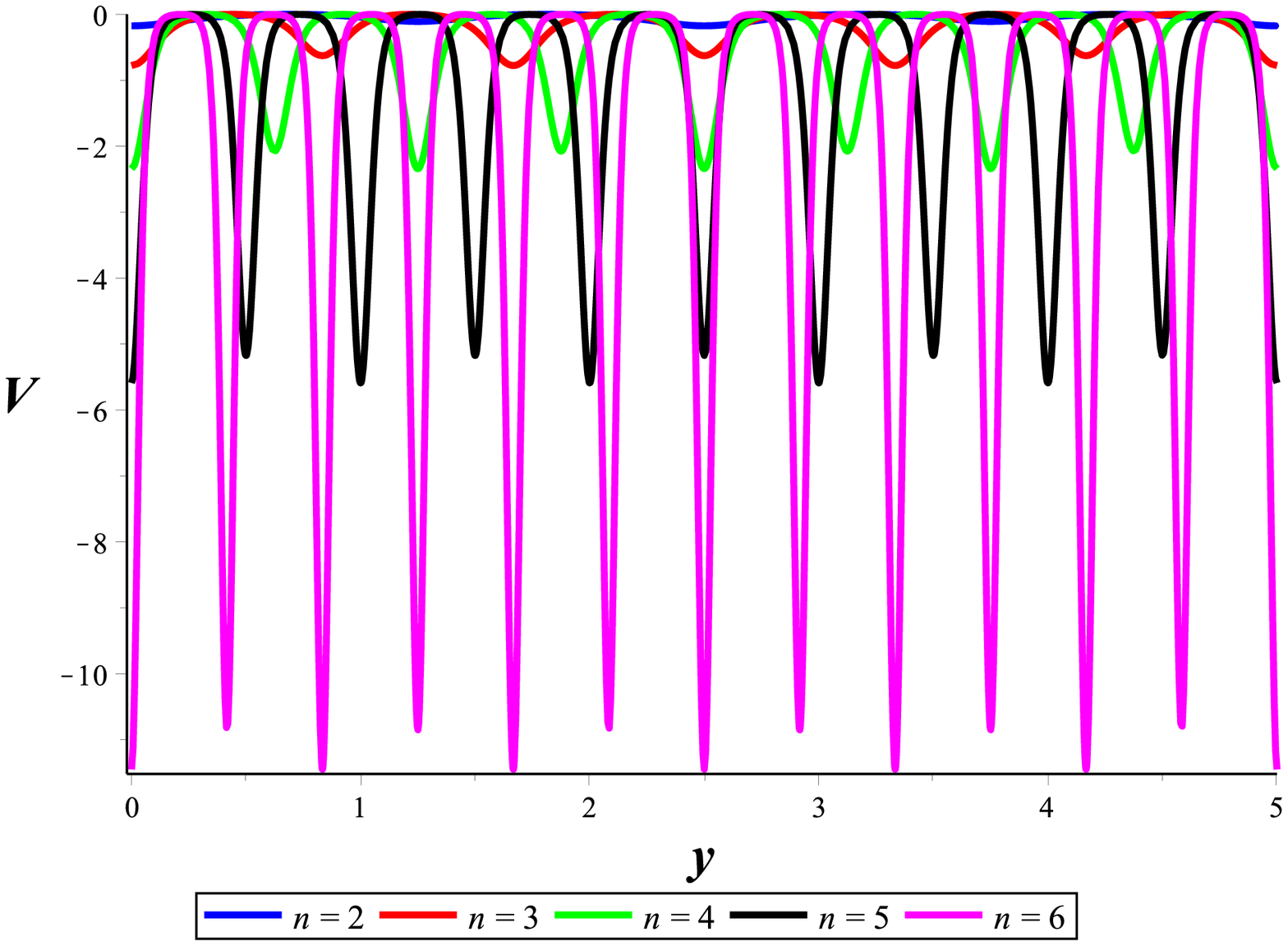}}
    \subfloat{\includegraphics[width=0.34\linewidth]{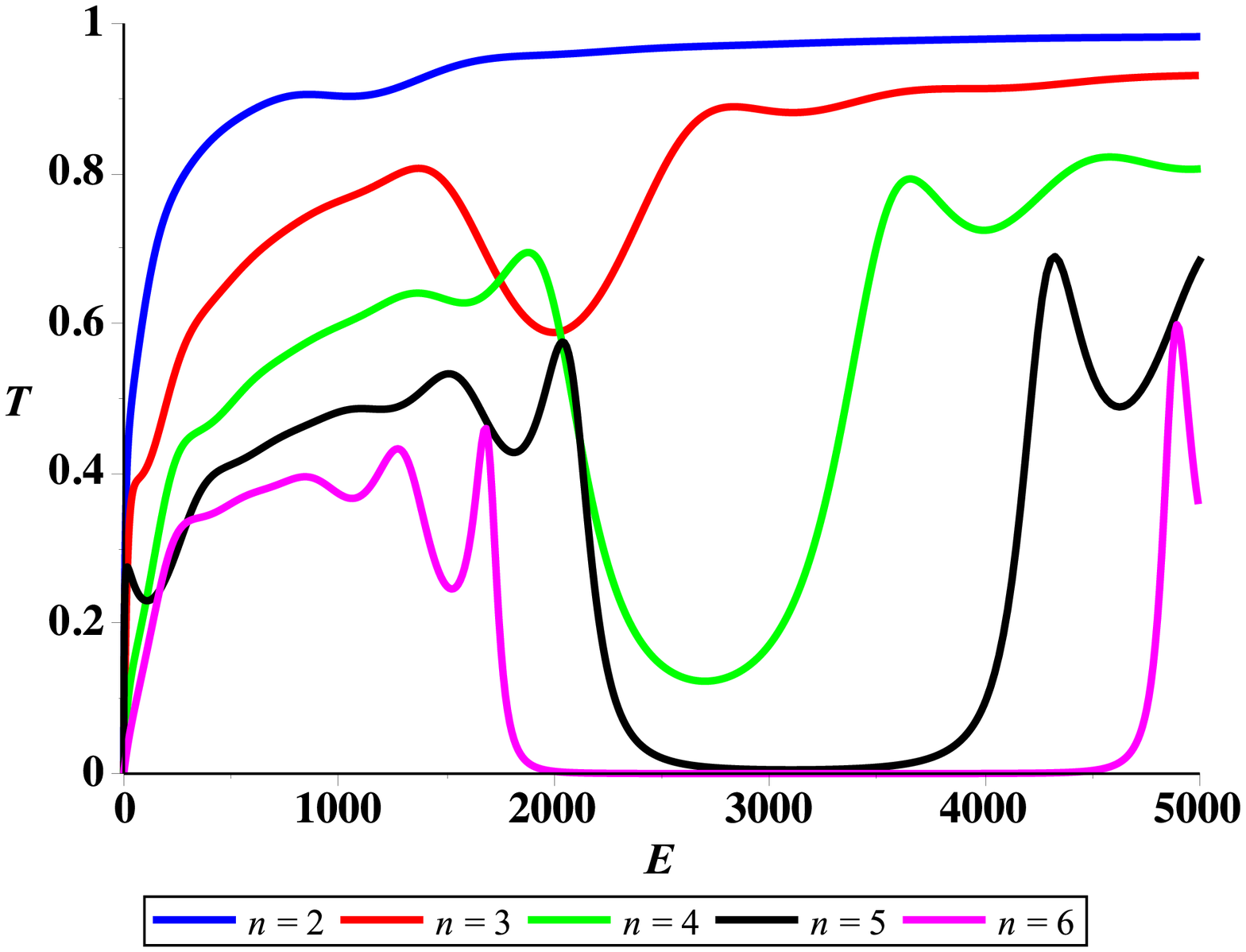}} 
          \caption{ (a) Wavy nanotube. (b) Geometric potential due to the deformation shown in (a) for different numbers of bumps. (c) Transmittance as function of incident energy (in meV). 
          } \label{4}
\end{figure*}

We look at three generic situations: a nanotube with a single bump, one with a pinch and a wavy structure. These are shown in  Figs.~\ref{Fig1}a, \ref{Fig2}a and \ref{4}a, respectively.
We create the corrugations by using the curve  
\begin{equation}
\rho(y)=R \pm \frac{\epsilon}{2} \left[1- \cos \left( 2\,{\frac {n\pi \,y}{L}} \label{curve}
 \right) \right]
\end{equation}
to generate the surface of revolution. In Eq. (\ref{curve}) the parameter $\epsilon$  regulates the strength of the deformation while $R$ gives the radius of the undistorted nanotube. The $+$ sign was used to generate the single bump and wavy structures and the $-$  sign for the pinched tube.  We used $n=1$ for the pinched and bumped nanotubes and $n=2\,,3\,,4\,,5\,,6$ for the wavy structures.

The geometric potential of a nanotube with a pinch and with a bump is seen in Figs.~\ref{Fig1}b and \ref{Fig2}b, respectively. Differently from the pinch, the bump has only positive Gaussian curvature. The result is a stronger geometric potential for the pinch which results in lower quasibound states and therefore  resonances in the transmittance shifted to lower energies. Figs.~\ref{Fig1}c and \ref{Fig2}c show the transmittance as  function of the incident energy. 
The pinched case, due to its deeper potential well, lowers the quasibound energy levels thus lowering the resonance peak positions as compared to the bumped case (see Supplementary Data for a direct comparison of the resonance peaks of the pinch and of the bump.).

In Fig. \ref{4} is shown the geometric potential and transmittance as function of incident energy for wavy structures with varying number of bumps. As the number of bumps increases, a 1D lattice on the tube starts to take form and the geometric potential starts to look like the Dirac comb \cite{Sakurai}, as seen in  the figure. The  periodicity  of the  potential minima  opens a gap  in the energy spectrum which becomes better defined with the increasing number of oscillations. The effect on the transmittance is seen in Fig. \ref{4}. As expected, the energy gap sensibly reduces the transmittance  and the effect becomes sharper with the number of bumps. For the curves displayed we fixed the tube length in 5 {\rm nm} and changed the number of bumps. Therefore the wavelength of the wavy perturbation changes with the number of bumps. Thus the width of the gap changes accordingly. The reader might ask why with a few bumps the gap is already so well defined. In order to answer this we evaluate the geometric potential at the position of its minima. There are two kinds: the deeper minima correspond to the pinched regions while the other ones correspond to the bumped regions.  The result for the deeper minima is
\begin{equation}
V_{\rm geo}={\frac {{\hbar}^{2} \left(L^2 -2\,{\pi }^{2}R \epsilon {n}^{2} \right) ^
{2}}{8{L}^{4}m{R}^{2}}}
\end{equation}
which means that the deepness of the minima grows with $n^4$. 
For the other minima, change $R \rightarrow R -\epsilon$. The dependence on $n$ is therefore the same. The half-width of the potential minima goes like $1/n$ since we are increasing the number of oscillations in a fixed nanotube and keeping its length $L$ fixed.  So, the potential quickly approaches a Dirac comb, giving the observed result. The periodic array of potential wells defines a Kroning-Penney look-alike resulting in the opening of bandgaps in the electronic structure. This is clearly seen in Fig.\ref{4}. As the number of wells increase the band gap becomes better defined.

\section{Conclusions}
The main object of this article is to illustrate how geometry can be used to manipulate electronic properties of nanostructures.  This was  used to study its effects on transport in three different examples involving nanotubes. The influence of perturbations of the cylindrical form of conducting nanotubes on their electronic transport properties was studied in this article with particular focus on the transmitivity. Curvature introduces a potential in the free particle Hamiltonian which is diagonalized numerically for chosen nanotube profiles. These include: a pinched nanotube, a tube with a bump and one with a sinusoidal deformation. In the first two cases we obtained results for different pinch/bump sizes. The results agree qualitatively with those of reference \cite{Taira20075270}. In the third case, for different  numbers of oscillations of the tube, the  energy gap becomes better defined with the increasing  number of oscillations. This suggests the possibility of using corrugated nanotubes as electronic filters.  In general, our results indicate the importance of curvature in the transport properties of ballistic electrons. The geometric approach used in this article can be easily generalized for different perturbations of the cylindrical geometry intrinsic to the nanotubes. Naturally, for a more realistic description of the transport properties, the geometric approach presented here must be extended to include interactions between  the charge carriers. This is presently under investigation and will be the subject of a follow-up article.

\ack{
F. S. and F. M. are thankful for the  warm hospitality at the Institut Jean Lamour of Universit\'e de Lorraine where part of this work was done. F. M. also thanks financial  support from the Institut Jean Lamour. This work has been partially supported by CNPq, CAPES and FACEPE (Brazilian agencies) and CNRS (France).}

\section*{References}
\bibliography{references}

\providecommand{\newblock}{}
\begin{thebibliography}{10}
\expandafter\ifx\csname url\endcsname\relax
  \def\url#1{{\tt #1}}\fi
\expandafter\ifx\csname urlprefix\endcsname\relax\def\urlprefix{URL }\fi
\providecommand{\eprint}[2][]{\url{#2}}

\bibitem{dacosta}
da~Costa R~C~T 1981 {\em Phys. Rev. A\/} {\bf 23}(4) 1982--1987

\bibitem{novoselov}
Meyer J~C, Geim A~K, Katsnelson M~I, Novoselov K~S, Booth T~J and Roth S 2007
  {\em Nature\/} {\bf 446} 60--63

\bibitem{PhysRevB.81.205409}
Atanasov V and Saxena A 2010 {\em Phys. Rev. B\/} {\bf 81}(20) 205409

\bibitem{PhysRevLett.84.1712}
Ozaki T, Iwasa Y and Mitani T 2000 {\em Phys. Rev. Lett.\/} {\bf 84}(8)
  1712--1715

\bibitem{doi:10.1021/nl061671j}
Yuzvinsky T~D, Mickelson W, Aloni S, Begtrup G~E, Kis A and Zettl A 2006 {\em
  Nano Letters\/} {\bf 6} 2718--2722

\bibitem{smith}
Smith B~W, Monthioux M and Luzzi D~E 1998 {\em Nature\/} {\bf 396} 323--324

\bibitem{:/content/aip/journal/jcp/142/19/10.1063/1.4921310}
Medina E, González-Arraga L~A, Finkelstein-Shapiro D, Berche B and Mujica V
  2015 {\em The Journal of Chemical Physics\/} {\bf 142} 194308

\bibitem{NanoRes7-1623}
Li Y, Ge J, Cai J, Zhang J, Lu W, Liu J and Chen L 2014 {\em Nano Research\/}
  {\bf 7} 1623--1630 ISSN 1998-0124

\bibitem{Ouyang27042001}
Ouyang M, Huang J~L, Cheung C~L and Lieber C~M 2001 {\em Science\/} {\bf 292}
  702--705

\bibitem{Marchi}
Marchi A, Reggiani S, Rudan M and Bertoni A 2005 {\em Phys. Rev. B\/} {\bf
  72}(3) 035403

\bibitem{Taira20075270}
Taira H and Shima H 2007 {\em Surface Science\/} {\bf 601} 5270 -- 5275 ISSN
  0039-6028 proceedings of the 10th ISSP International Symposium on Nanoscience
  at Surfaces

\bibitem{PhysRevLett.78.1932}
Kane C~L and Mele E~J 1997 {\em Phys. Rev. Lett.\/} {\bf 78}(10) 1932--1935

\bibitem{PhysRevLett.79.5082}
Egger R and Gogolin A~O 1997 {\em Phys. Rev. Lett.\/} {\bf 79}(25) 5082--5085

\bibitem{nature397-598}
Bockrath M, Cobden D~H, Lu J, Rinzler A~G, Smalley R~E, Balents L and McEuen
  P~L 1999 {\em Nature\/} {\bf 397} 598--601

\bibitem{PhysRevLett.79.5086}
Kane C, Balents L and Fisher M~P~A 1997 {\em Phys. Rev. Lett.\/} {\bf 79}(25)
  5086--5089

\bibitem{PhysRevLett.113.227205}
Chang C~H, van~den Brink J and Ortix C 2014 {\em Phys. Rev. Lett.\/} {\bf
  113}(22) 227205

\bibitem{PhysRevLett.112.257203}
Gaididei Y, Kravchuk V~P and Sheka D~D 2014 {\em Phys. Rev. Lett.\/} {\bf
  112}(25) 257203

\bibitem{doi:10.1142/S2010324713400067}
Chen K~C and Chang C~R 2013 {\em SPIN\/} {\bf 03} 1340006

\bibitem{0953-8984-23-17-175301}
Atanasov V and Saxena A 2011 {\em Journal of Physics: Condensed Matter\/} {\bf
  23} 175301

\bibitem{0295-5075-98-2-27001}
Onoe J, Ito T, Shima H, Yoshioka H and ichi Kimura S 2012 {\em EPL (Europhysics
  Letters)\/} {\bf 98} 27001

\bibitem{lent}
Lent C~S and Kirkner D~J 1990 {\em Journal of Applied Physics\/} {\bf 67}
  6353--6359

\bibitem{Sakurai}
Sakurai J~J 1993 {\em {Modern Quantum Mechanics (Revised Edition)}\/} 1st ed
  (Addison Wesley) ISBN 0201539292

\end{thebibliography}


\providecommand{\newblock}{}
\begin{thebibliography}{1}
\expandafter\ifx\csname url\endcsname\relax
  \def\url#1{{\tt #1}}\fi
\expandafter\ifx\csname urlprefix\endcsname\relax\def\urlprefix{URL }\fi
\providecommand{\eprint}[2][]{\url{#2}}

\bibitem{dacosta}
da~Costa R~C~T 1981 {\em Phys. Rev. A\/} {\bf 23}(4) 1982--1987

\bibitem{Marchi}
Marchi A, Reggiani S, Rudan M and Bertoni A 2005 {\em Phys. Rev. B\/} {\bf
  72}(3) 035403

\end{thebibliography}

\end{document}


\begin{textblock*}{100mm}(\textwidth,-2cm)
\textcolor{blue}{\large Supplementary Data}
\end{textblock*}

\title{QM in curved surfaces, consistency check and MAPLE code}

\author{Fernando Santos
\footnote{On leave from: Departamento de Matem\'atica  Universidade Federal 
de Pernambuco, 50670-901, Recife, PE, Brazil}}
\address{Wolfson Centre for Mathematical Biology, Mathematical Institute, University of Oxford, OX1 3LB Oxford, U.K.}
\author{S\'ebastien Fumeron, Bertrand Berche}
\address{ Statistical Physics Group, IJL, UMR Universit\'e de Lorraine - CNRS 7198
 BP 70239,  54506 Vand\oe uvre les Nancy, France}
 \author{Fernando Moraes
\footnote{On leave from: Departamento de F\'\i sica, CCEN, Universidade Federal da Para\'\i ba, Caixa Postal 5008, 58051-900, Jo\~ao Pessoa, PB, Brazil }}

\address{Departamento de F\'\i sica and Departamento de Matem\'atica, Universidade
Federal de Pernambuco, 
 50670-901 Recife, PE, Brazil}
 
\date{\today}

\begin{abstract}
Supplementary Data includes a check of consistency of our results by comparison with the relevant literature and also provides the access for the community to the MAPLE code which enables one to proceed to further calculations in cylindrical geometries. 
\end{abstract}



\maketitle

\section{Schr\"odinger equation in a curved geometry}
The fact that real quasi-two-dimensional systems are embedded in Euclidean three-dimensional space and have a finite but very small thickness has important consequences for the dynamics of the particle. First, the kinetic energy operator will be affected since the Laplacian has to follow the local geometry of the surface. This is clearly seen by considering the 3D Laplace operator in spherical coordinates 
\begin{eqnarray}\nabla^2&=&\frac{1}{r^2}\frac{\partial}{\partial r}\left(r^2 \frac{\partial}{\partial r}\right) 
\nonumber\\ &&+ \frac{1}{r^2 \sin^2 \phi}\frac{\partial^2}{\partial^2 \theta^2}+ \frac{1}{r^2 \sin \phi}\frac{\partial}{\partial \phi}\left(\sin\phi\frac{\partial}{\partial \phi}\right).\end{eqnarray} 
Since the wavefunction of a quantum particle moving on the surface of the sphere does not depend on the radial coordinate, the kinetic energy operator for such particle is 
\begin{equation}-\frac{\hbar^2}{2m}\left[ \frac{1}{r^2 \sin^2 \phi}\frac{\partial^2}{\partial^2 \theta^2}+ \frac{1}{r^2 \sin \phi}\frac{\partial}{\partial \phi}\left(\sin\phi\frac{\partial}{\partial \phi}\right) \right].\end{equation} 
The term between square brackets is the Laplace-Beltrami operator.
Considering a finite thickness $d$ for the ``surface" and a confining potential given by an infinite square well in the normal direction, da Costa found that the Schr\"odinger operator for a free particle in this confined geometry, in the limit $d\rightarrow 0$, is \cite{dacosta}
\begin{equation}
H = -\frac{\hbar^2}{2m}\triangle_{LB} + V_{\rm geo}, \label{HdaCosta}
\end{equation} 
where the Laplace-Beltrami operator is obtained from the curvilinear coordinates $x^i$ intrinsic to the surface and the metric tensor $g_{ij}$ ($g$ its determinant)
\begin{equation}
\triangle_{LB} = \sum_{i,j}\frac{1}{\sqrt{g}}\frac{\partial}{\partial x^i}\left(\sqrt{|g|}g^{ij}\frac{\partial}{\partial x^j}  \right), \label{LB}
\end{equation}
and the geometric potential is given by
\begin{equation}
V_{\rm geo}=- \frac{\hbar^2}{2m}\left(M^2 - K \right). \label{geompotential}
\end{equation}
Here, $M$ and $K$ refer to the mean and Gaussian curvatures, respectively. The geometric potential is a direct consequence of the quantization of the motion normal to the surface. 
This is why it does not appear in the 2D sphere example given above.

\section{Consistency check}

\begin{figure}[h!]
 
  \centering
    \includegraphics[width=0.35\textwidth]{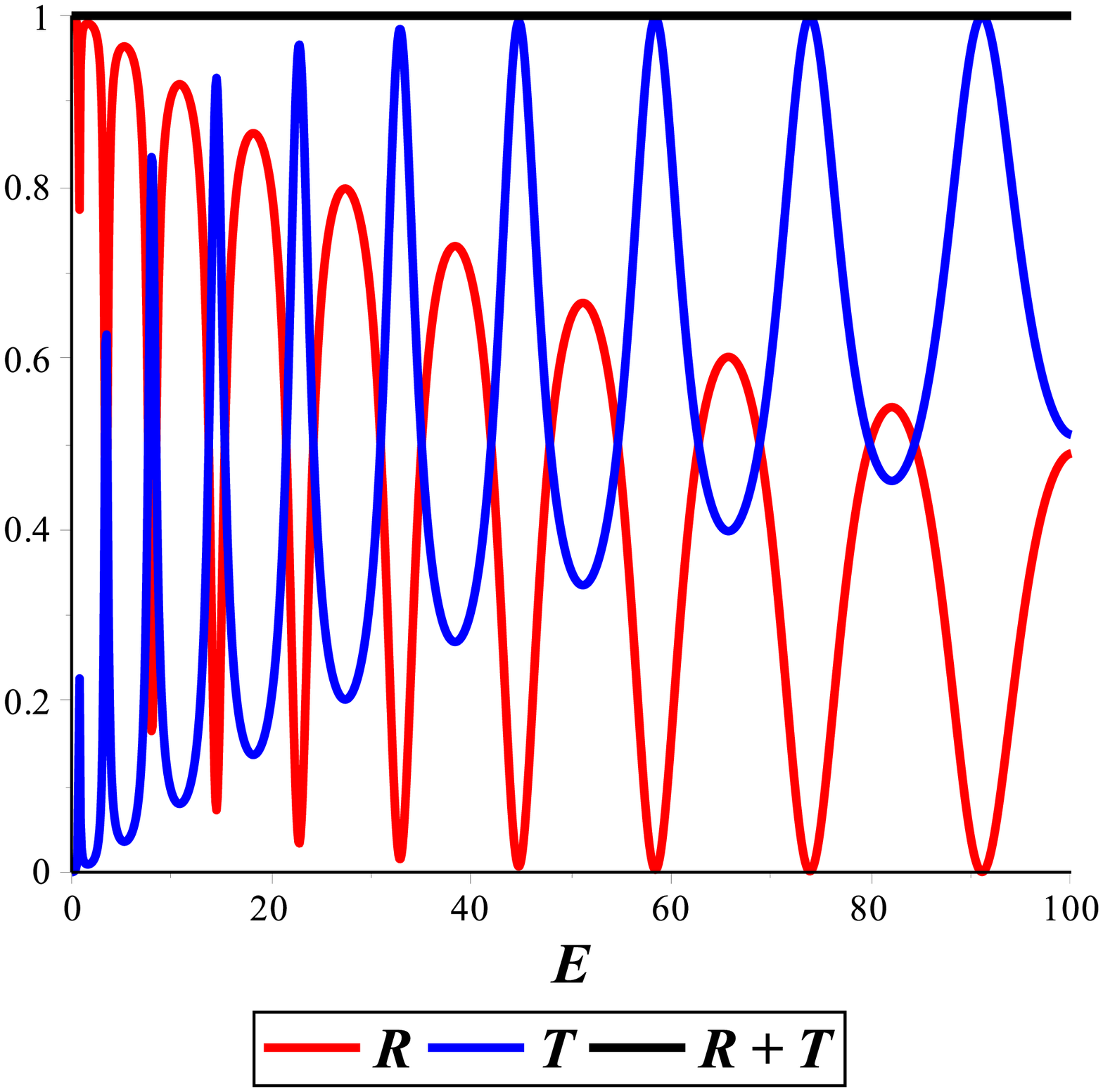} 
     \caption{Transmittance T and reflectance R as  function of injection energy E for the symmetric junction studied in reference \cite{Marchi}. }\label{figmarchi}
\end{figure}

In order to check our numerical code we reproduced the transmittance $T$ versus injection energy $E$ plot for the axially symmetric cylindrical junction   of reference \cite{Marchi}.  As in that work, we used $\ell=0$, $R_2=0.397$ nm for the cylinder radius in the region $y>L$  and the electron effective mass $m^* = 0.173 m_e$ constant throughout the junction. The result for $R_1=50$ nm (radius in the region $y<0$) is  plotted in Fig. \ref{figmarchi} which is in  perfect agreement with Fig. 7 of reference \cite{Marchi}. The consistency of our numerical calculations was evaluated by checking if the output of Eqs. 
\begin{equation}
T=\frac{k_L}{k_0} \left| \varphi(L) \right|^2  \label{transnew}
\end{equation}
and
\begin{equation}
R=|\varphi(0)-1 |^2 \label{refnew} .
\end{equation}
yields $T+R=1$ verifying the conservation of probability density. This is also seen in Fig. \ref{figmarchi} where the reflectance $R$ as a function of the injection energy and  the sum $R+T$ are plotted. Clearly, the results are consistent.

\section{Pinched and bumped nanotube}

\begin{figure}[h!]
 
  \centering
    \includegraphics[width=0.35\textwidth]{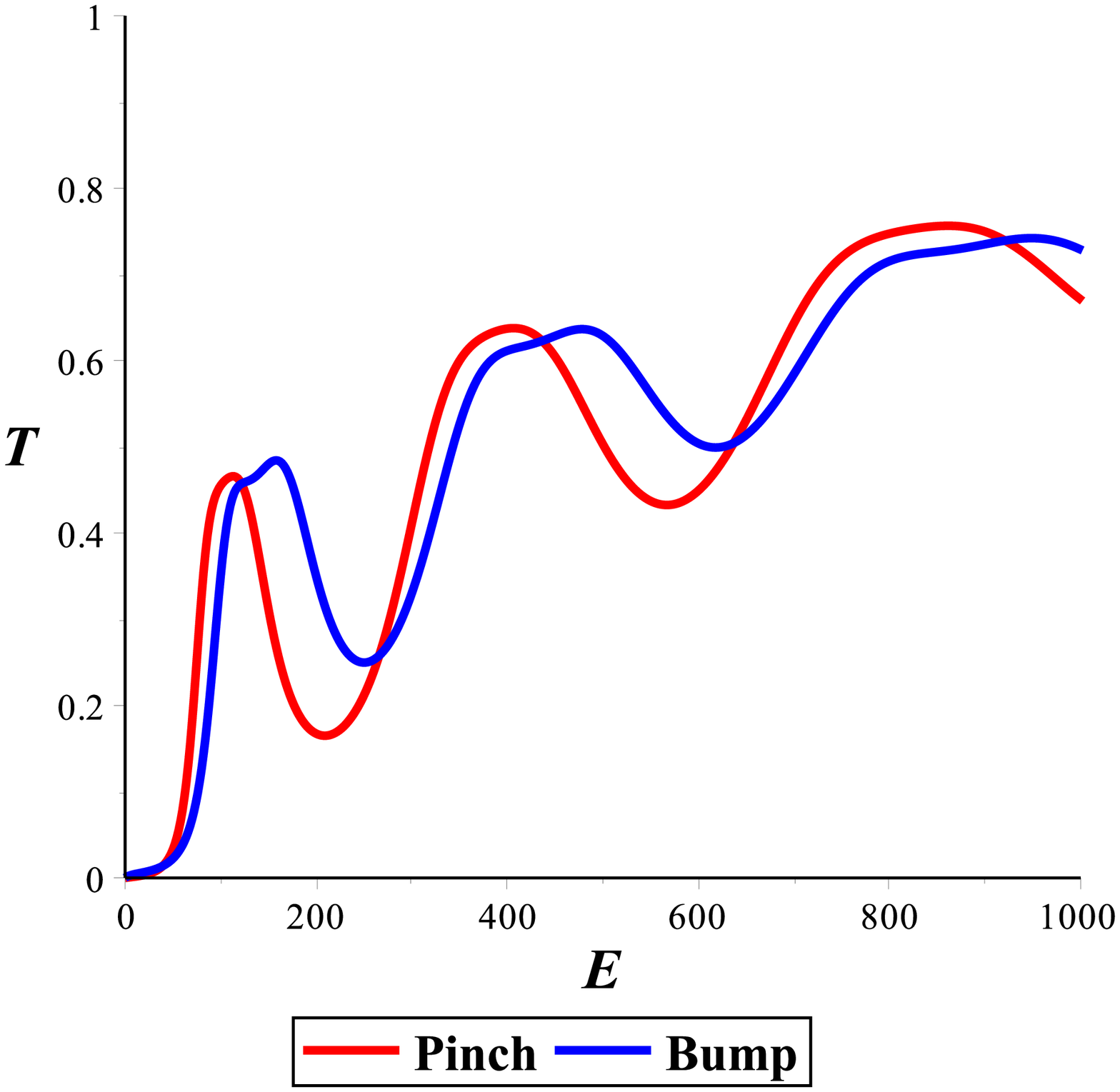} 
     \caption{Transmittance T and reflectance R as  function of injection energy E for $\epsilon=0.95$.}\label{fig5}
\end{figure}

The shape of the transmittance vs energy for pinched and bumped nanotubes seem to be very similar, but the potential wells are deeper in the pinched case. 
To make the shift to lower energy in the pinched case clearer we plot in Fig. \ref{fig5} the transmittance as function of incident energy for both the bumped and pinched nanotubes for $\epsilon =0.95$. 

%

\section*{References}

\bibliography{references}

\section*{APPENDIX: MAPLE code}

\includepdf[pages= - ]{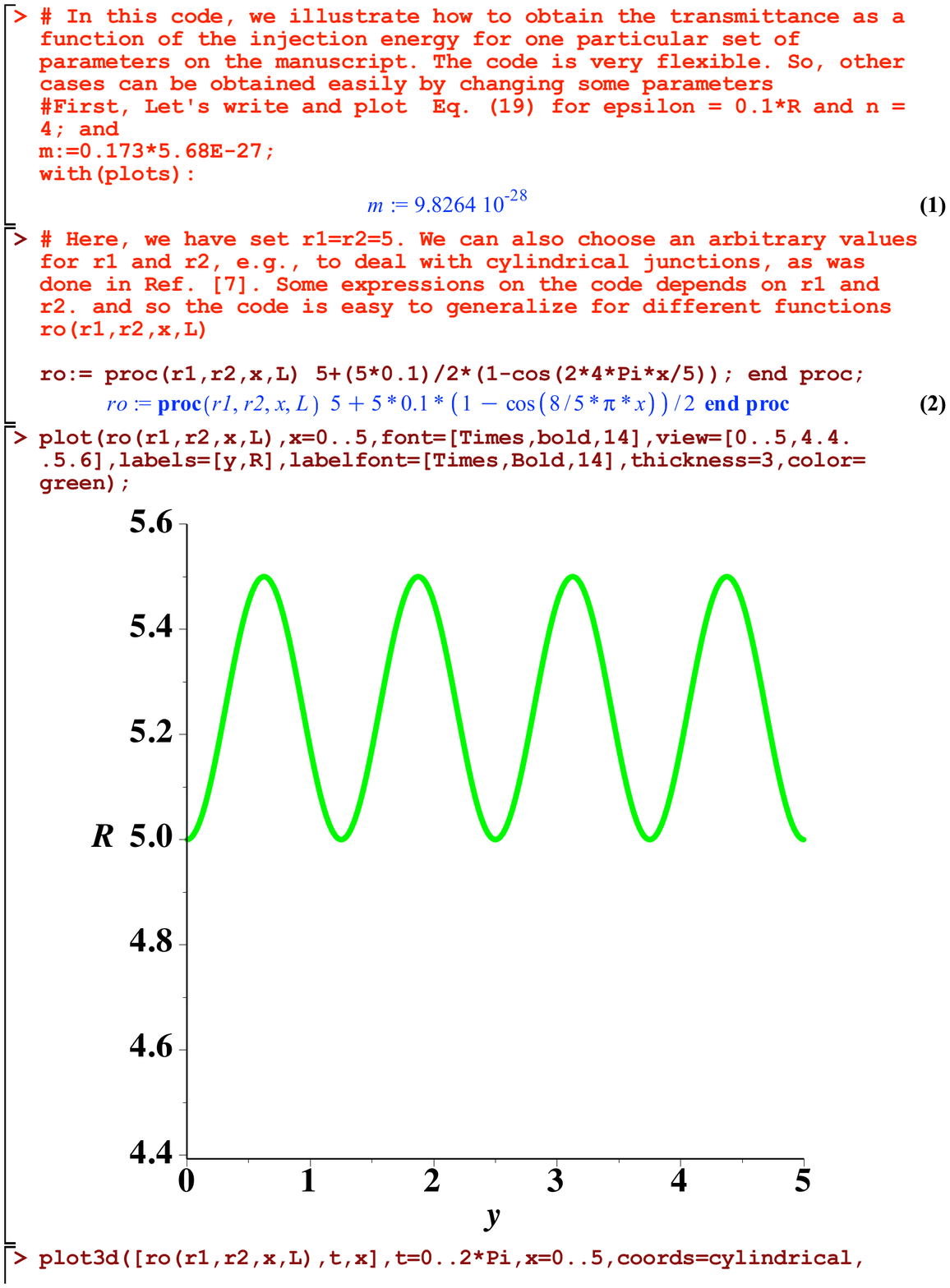}